\documentclass[12pt,letterpaper]{article}
\pdfoutput=1
\usepackage[left=2cm,top=1cm,right=3cm,nohead]{geometry}
\usepackage[dvipsnames]{xcolor}
\usepackage{colortbl}
\usepackage[utf8]{inputenc} 
\usepackage{color}
\usepackage{epsfig,graphics,latexsym,amsfonts,mathtools,amsthm,amssymb,amsbsy,multirow,slashed,mathrsfs,wasysym,textcomp,subfigure,wrapfig,comment,bbold,array,longtable,multirow,setspace,amsmath,pifont, enumitem}
\usepackage{tabularx}
\usepackage{graphicx}
\usepackage{graphbox}
\usepackage{setspace}
\usepackage{multirow}
\usepackage{rotate}	
\usepackage{slashed}
\usepackage{accents}
\usepackage[bf]{caption}
\usepackage{braket}
\usepackage{fancyhdr}
\usepackage{comment}
\usepackage{float}
\usepackage{tikz}
\usetikzlibrary{mindmap,trees,shadows}
\colorlet{color1}{gray!25}
\usetikzlibrary{positioning}
\usetikzlibrary{intersections} 

\newlength{\PicScale}
\usepackage[UKenglish]{babel}
\usepackage[toc,page]{appendix}
\usepackage{hhline}
\usepackage{geometry}
\usepackage{cite}
\usepackage{datetime}
\usepackage{bbm} 
\usepackage{bm} 
\usepackage[colorlinks=true,
linkcolor=RoyalPurple, 
citecolor=RoyalPurple,
filecolor=RoyalPurple,
urlcolor=RoyalPurple,
linktoc=page, 
pdfstartview=FitV,
bookmarksopen=true]{hyperref}
\hypersetup{
	pdftitle={},
	pdfauthor={},
	pdfsubject={}
} 
\usepackage{cleveref}
\usepackage{subfiles}
\definecolor{Gray}{gray}{0.94}
\usepackage{simpler-wick}

\newlength{\dhatheight}

\numberwithin{equation}{section}

\title{CHL decompactification}

\usepackage{xcolor}

\def\Vnote#1{{\textcolor{Purple}{[V: #1]}}}

\DeclareFontFamily{U}{rsf}{}
\DeclareFontShape{U}{rsf}{m}{n}{
  <5> <6> rsfs5 <7> <8> <9> rsfs7 <10-> rsfs10}{}
\DeclareMathAlphabet\Scr{U}{rsf}{m}{n}

\def\rep#1{{{\boldsymbol{#1}}}}

\def\Z{{\mathbb Z}}

\def\Tr{\operatorname{Tr}}

\def\GO{\operatorname{O{}}}

\def\GE{\operatorname{E}}

\def\su{\operatorname{\mathfrak{su}}}
\def\Lu{\operatorname{\mathfrak{u}}}
\def\Le{\operatorname{\mathfrak{e}}}
\def\Lg{\operatorname{\mathfrak{g}}}

\def\la{\langle}
\def\ra{\rangle}

\def\ff#1#2{{\textstyle\frac{#1}{#2}}}

\def\cC{{\cal C}}

\def\cF{{\cal F}}
\def\cG{{\cal G}}
\def\cH{{\cal H}}

\def\cJ{{\cal J}}
\def\cK{{\cal K}}

\def\cX{{\cal X}}

\def\cGt{{\widetilde{\cal G}}}
\def\cFt{{\widetilde{\cal F}}}

\def\fabc{{f^{ab}_{~~~c}}}
\def\fabd{{f^{ab}_{~~~d}}}

\def\ep{{\epsilon}}

\newcommand\taub{\overline{\tau}}

\newcommand\ft{\widetilde{f}}

\newcommand\Zt{\widetilde{Z}}

\begin{document}

\numberwithin{equation}{section}

\begin{flushright}
\end{flushright}
\pagestyle{empty}

\vspace{20mm}

 \begin{center}
{\LARGE \textbf{{A twist at infinite distance in the CHL string}}}

\vspace{10mm} {\Large Veronica Collazuol$^a$ and Ilarion V. Melnikov$^b$ }\\

\vspace{5mm}

$^a$Institut de Physique Th\'eorique,
Universit\'e Paris Saclay, CEA, CNRS, Orme des Merisiers, \\ F-91191 Gif sur Yvette, France \\
$^b$Department of Physics and Astronomy, James Madison University, Harrisonburg, VA 22807,\\ USA

\vspace{.5mm} {\small\upshape\ttfamily veronica.collazuol@ipht.fr, melnikix@jmu.edu} \\

\vspace{15mm}

\textbf{Abstract}
\end{center}
We analyze a space-time algebra of BPS states that emerges in the infinite distance limit in the moduli space of the nine-dimensional CHL string as the theory decompactifies to the ten-dimensional $\text{E}_8 \times \text{E}_8$ heterotic string. We find an affine algebra as expected from the heterotic case, but in a twisted version.

\vspace{3mm}

\newpage
\setcounter{page}{1}
	\pagestyle{plain}
		
\tableofcontents
\newpage

\section{Introduction}
As the parameters of a physical system are tuned to approach a singular limiting point, the system often exhibits remarkable enhanced symmetries and patterns of universal behavior.  Recently there has been interest and progress in understanding a class of such infinite-distance limiting points in the moduli space of string compactifications and more generally in the moduli space of any putative theory of quantum gravity.   Reviews of this work in light of the Swampland Program \cite{Vafa:2005ui} may be found in~\cite{Brennan:2017rbf,Palti:2019pca,Agmon:2022thq,vanBeest:2021lhn}.
%
For example, the Swampland Distance Conjecture \cite{Ooguri:2006in} predicts that, independently of the nature of the compactification,  an infinite tower of states become massless exponentially in the geodesic distance as one approaches these regions. This tower is conjectured to be describable in some duality frame as either a tower of Kaluza-Klein (KK) modes or a tower of oscillators of a critical string becoming tensionless \cite{Lee:2019wij}.

In the presence of perturbative gauge sectors, it is known that when these infinite distance points correspond to decompactification limits in some duality frame, the tower of massless states enhances the symmetry algebra to its affine version, as has been proven in the case of heterotic theories on tori \cite{Collazuol:2022jiy,Collazuol:2022oey} and in the dual case of F theory on K3 \cite{Lee:2021qkx,Lee:2021usk} \footnote{See also \cite{Cvetic:2022uuu} for equivalent results obtained from the string junction perspective.}. In particular, if we start from a theory in $d$ dimensions and decompactify to a theory in $d+k$ dimensions with gauge symmetry $G$, then from the lower dimensional point of view we obtain an affine version of $G$ with $k$ central extensions. In the heterotic frame the central extensions correspond to the bosons associated to the decompactified directions, and the loop part of the algebra is built from the KK modes of the horizontal algebra $G$ (which have been classified in heterotic compactifications with 16 supercharges from 9 to 6 dimensions in a series of papers \cite{Fraiman:2018ebo,Font:2020rsk,Fraiman:2021soq,Fraiman:2022aik}).

In this paper, we focus on a slightly more complicated case: the nine-dimensional CHL string \cite{Chaudhuri:1995fk}, which can be obtained as an orbifold of the $\text{E}_8 \times \text{E}_8$ heterotic theory on $S^1$, whose effect is to reduce the rank of the gauge symmetry from 17 to 9. As was argued in \cite{Chaudhuri:1995bf}, in decompactifying the circle direction one recovers the Cartan generators that were projected out by the orbifold, obtaining back the ten dimensional $\text{E}_8 \times \text{E}_8$ heterotic theory. Thus, to this extent, the decompactification limit in the CHL string is the same as the one of the $\text{E}_8 \times \text{E}_8$ heterotic string on $S^1$. We show that from the nine-dimensional point of view there is an important difference. The algebra in the heterotic theory enhances to $(\Le_9 \oplus \Le_9) / \sim$ due to the presence of the KK tower, where the $\sim$ stands for the identification of the central extensions of the two $\Le_9$ factors. In the case at hand the momentum towers arrange in a twisted version of the affine $(\Le_9 \oplus \Le_9) / \sim$ , while the central extension is still identified between the two $\Le_9$ factors. 

This gives an interesting prediction for dual theories. For example, if we further compactify the nine dimensional CHL on a circle, the theory is believed to be dual to an F-theory compactification on a K3 with frozen singularities \cite{Witten:1997bs}. Our result predicts the appearance of a doubly affine twisted algebra in the limit dual to the full decompactification of the eight dimensional CHL string, and understanding this in the M/F-duality frame could provide insight into the physics of frozen singularities. 

This note is organized as follows. In section \ref{sec:9dchl} we review the construction of the CHL string in nine dimensions; in section \ref{sec:rinfty} we describe the decompactification limit $R \to \infty$ and how the twisted version of the BPS algebra $(\Le_9 \oplus \Le_9) / \sim$ is realized in terms of asymptotically conserved holomorphic currents, and in section \ref{sec:rzero} we study how this algebra arises in the T-dual frame, where $R \to 0$.  Finally, we give an overview in Section \ref{sec:conclusion}.

\section{The nine-dimensional CHL string}
\label{sec:9dchl}
In this section we recall the main features of the CHL string that we will use in the following to characterize the current algebra arising in the decompactification limit. 
\subsection{Orbifold construction of the CHL string}
In the construction of \cite{Chaudhuri:1995bf}, which is most suited to our analysis, the CHL theory \cite{Chaudhuri:1995fk} in nine dimensions is obtained as a $\mathbb{Z}_2$ orbifold of the $\text{E}_8 \times \text{E}_8$ heterotic string theory compactified on $S^1$.  For our purposes it is sufficient to work with the Narain sector of the internal CFT, a theory with central charges $c=17$ and $\bar{c} = 1$, which is described as follows.
Let $X^I(z)$ ($I=1,...,16$) be the holomorphic compact chiral bosons on the heterotic torus and $X^9(z,\bar{z})$ the non-chiral boson for the circle of radius $R$, normalized as
\begin{align}
    \label{eqn:oldII}
        X^{I}(z) X^{J}(w) & \sim - \delta^{IJ} \log(z-w) \, , \\
    \label{eqn:99ope}
        X^9(z,\bar{z}) X^9(w,\bar{w}) & \sim -  \frac{1}{2R^2} \log|z-w|^2 \, .  
\end{align}
We decompose the non-chiral boson as
\begin{align}
    X^9(z,\bar{z}) = \frac{1}{\sqrt{2}} \left(X^9_L(z)+X^9_R(\bar{z})\right) \, .
\end{align}
We split the Cartan indices of the heterotic bosons between the two $\text{E}_8$ factors as $I=(i,i+8)$, with $i=1,..,8$, and denote $X^i_1(z) = X^i(z)$ and $X^i_2(z)=X^{i+8}(z)$, so that the subscripts $1$ and $2$ signify the copy of $\text{E}_8$.
The orbifold action consists of the exchange $\mathcal{E}$ of the two $\text{E}_8$ factors, together with a half-shift $\mathcal{S}$ in the compact circle direction: 
\begin{align}
    \mathcal{E} & :  \quad X^{i}_1(z) \leftrightarrow X^{i}_2(z) \, , \\
    \mathcal{S} & : \quad X^9(z,\bar{z}) \to X^9(z,\bar{z}) + \pi \,.
\end{align}
This is a symmetry of the heterotic $\text{E}_8 \times \text{E}_8$ theory on $S^1$ in the subspace of its moduli space 
\begin{equation}
    \mathcal{M}_{\text{het}} = \frac{O(1,17; \mathbb{R})}{O(17;\mathbb{R})} \Big/ O(1,17; \mathbb{Z}) \, ,
\end{equation}
characterized by a symmetric Wilson line $A=(a,a)$.\footnote{Here $a$ is valued in the torus $\mathbb{R}^8/\Gamma_8$, where $\Gamma_8$ is the $\text{E}_8$ lattice.} In 9 dimensions this is a freely acting orbifold, whose effect is to reduce the rank of the gauge group from 17 to 9, while not breaking any supersymmetry.  The primary states in the Narain CFT are labelled by integer momentum $\text{n}$ and winding $\text{w}$ along the $S^1$ and by heterotic momenta $\pi=(\pi_1, \pi_2)$, and the orbifold acts on these as
\begin{equation}
    \mathcal{E}\mathcal{S} \ket{\text{n},\text{w},\pi_1,\pi_2} = (-1)^{\text{n}} \ket{\text{n},\text{w},\pi_2,\pi_1} \, .
\end{equation}

The Hilbert space of the orbifold theory has the untwisted sector and a single twisted sector.  In order to describe these spaces, it is useful to change the basis for the heterotic bosons from $\{X^i_1(z),X^{i}_2(z) \}$ to the symmetric and antisymmetric combinations
\begin{equation}
    X^i_{\pm}(z) = \frac{1}{\sqrt{2}} \left(X^{i}_1(z) \pm X^{i}_2(z)\right) \xrightarrow[]{\mathcal{E}\mathcal{S} } \pm X^i_{\pm}(z) \, .
\end{equation}
The states belonging to the untwisted sector are constructed from bosons with untwisted boundary conditions
\begin{align}
    X^{i}_{\pm}(e^{2\pi i}z) = X^{i}_{\pm}(z)+\frac{1}{\sqrt{2}} Q^i_{\pm} \, , \\
        X^{9}(e^{2\pi i}z,e^{2\pi i}\bar{z}) = X^{9}(z,\bar{z})+2\pi \text{w} \, ,
\end{align}
where $Q_{\pm} \in \Gamma_8$ are fixed vectors in the $\text{E}_8$ root lattice, and $\text{w} \in \mathbb{Z}$ is the winding number; in the Hilbert space of the Narain theory there are sectors labelled by each allowed choice of $\{ Q_{\pm}, \text{w} \}$.
In the sector twisted with respect to the $\mathbb{Z}_2$ orbifold action, closed strings have boundary conditions which are additionally twisted with respect to the $\mathcal{E} \mathcal{S}$ transformation 
\begin{align}
    X^{i}_{\pm}(e^{2\pi i}z) & = \pm X^{i}_{\pm}(z)+\frac{1}{\sqrt{2}} Q'^i_{\pm} \, , \\
        X^{9}(e^{2\pi i}z,e^{2\pi i}\bar{z}) & = X^{9}(z,\bar{z})+\pi+2\pi \text{m} = X^{9}(z,\bar{z})+2\pi \text{w}\, ,
\end{align}
where $Q'_{\pm} \in \Gamma_8$ and $\text{m}\in \mathbb{Z}$, so that  $\text{w} \in \mathbb{Z}+\frac{1}{2}$. Being antisymmetric under the orbifold, in the twisted sector $X_-^i(z)$ has half-integer mode expansion 
\begin{equation}
    X_-^i(z)= \frac{1}{2\sqrt{2}}Q^i_- + \sqrt{\frac{1}{2}} \sum_{r \in \mathbb{Z}+\frac{1}{2}} \frac{\alpha_{-;r}^i}{r z^{r}} \, .
\end{equation}
The moduli space of the CHL theory is given by
\begin{equation}
    \mathcal{M}_{\text{CHL}} = \frac{O(1,9; \mathbb{R})}{O(9;\mathbb{R})} \Big/ O(1,9; \mathbb{Z}) \, ,
\end{equation}
and it is parametrised by the radius $R$ and Wilson line $a$. $O(1,9; \mathbb{Z})$ is the T-duality group, and it corresponds to the group of automorphisms of the lattice $II_{1,9}$ spanned by the charge vectors of the physical states $Z=\ket{\text{n},2\text{w},\rho}$. A particularly interesting duality transformation is inherited from the radius inversion $R \to \frac{1}{R}$ duality of the Narain CFT, which in the CHL case interchanges states between the two sectors of the CHL orbifold. Its action on the moduli and charges is 
\begin{align}
    R^2 + a^2 & \to \frac{4}{R^2 + a^2} \, , \nonumber\\
    a & \to \frac{2a}{R^2 + a^2} \, , \nonumber \\
    2\text{w} & \leftrightarrow \text{n} \, , \nonumber \\
    \label{eqn:a}
    \rho & \to - \rho~,
\end{align}
which implies that untwisted states with even momentum are mapped to untwisted states ($\text{w} \in \mathbb{Z}$), while the ones with odd momentum are dual to twisted states ($\text{w} \in \mathbb{Z}+\frac{1}{2}$). 

\subsection{Symmetry enhancements}
In supersymmetric perturbative heterotic compactifications, space-time gauge bosons arise in both the right (supersymmetric) and left moving sectors of the string. The former are universal, namely they are present at each point in moduli space, and give a $\Lu(1)^d$ anti-holomorphic current algebra in the case of $T^d$ compactifications. The latter are in one-to-one correspondence with holomorphic currents, which depend on the point of moduli space. In this paper, we will be concerned with the second set.  
At a generic point $\{R, a\} $, they consist of
\begin{itemize}
    \item the untwisted current 
        \begin{equation}
        \label{eqn:circle}
            J^9(z) = i \partial X^9(z) \, ,
        \end{equation}
    corresponding to the state generated by the left oscillator $\alpha^9_{-1}\ket{0}$,    
    \item the 8 untwisted currents 
        \begin{equation}
        \label{eqn:cart}
             J^{i}_+(z) = i (\partial X^i_1(z)+\partial X^i_2(z))\, ,
        \end{equation}
    corresponding to the untwisted states $\frac{1}{\sqrt{2}}(\alpha^i_{-1}+\alpha^{i+8}_{-1}) \ket{0}$~,    
\end{itemize}
which together build the algebra $\Lu(1)^9$ in the holomorphic sector of the theory. At particular loci in the moduli space there appear additional conserved currents $J^{\alpha}(z)$ that can belong either to the untwisted or  twisted sector, corresponding to primary states $ \ket{\text{n},\text{w},\rho}$ with weights $\text{h}=1, \, \bar{\text{h}}=0$. 
In nine dimensions, the modes of the currents $\{ J^{\alpha}(z) \} $ include the ladder operators associated to the roots $\{ \alpha \}$ of a semisimple ADE algebra $\Lg_r$ \footnote{For a full classification of the possible $\Lg_r$ in 9 dimensions (as well as the $r=10$ case in 8 dimensions), see \cite{Font:2021uyw}.} of rank $r\leq 9$ and give the symmetry enhancement pattern
\begin{equation}
    \Lu(1)^9 \to (\Lg_r)_2 \times \Lu(1)^{9-r}~,
\end{equation}
where the subscript 2 denotes the level of the algebra. 

As an example, let us consider the locus in moduli space given by $a=0$ and generic $R$. Here the theory displays the symmetry $(\Le_8)_2 \oplus \Lu(1)$.\footnote{At $R=\sqrt{2}$ the $\Lu(1)$ factor is further enhanced to $\su(2)$.} 
The $\Lu(1)$ current is in \eqref{eqn:circle}; the Cartan generators of $(\Le_8)_2$ are given by \eqref{eqn:cart}, and the ladder operators are
the symmetric combinations of the $(\Le_8)_1 \oplus (\Le_8)_1$ ones (which have positive eigenvalue under the $\mathcal{E}\mathcal{S}$ action, and so are kept by the orbifold projection):
\begin{equation}
\label{eqn:currents}
    J^{\alpha}_+(z) = c_{\alpha} (:e^{i \alpha^i X^{i}_1(z)}:+:e^{i \alpha^i X^{i}_2(z)}:) \,, \, \alpha \in \Gamma_8 \, , 
\end{equation}
where $c_{\alpha}$ is the cocycle factor ensuring the correct commutation properties of the currents. Denoting by $a$ the adjoint representation of $\Le_8$, \eqref{eqn:cart} and \eqref{eqn:currents} together satisfy the level $2$ current algebra
\begin{align}
\label{eqn:jpjp}
    J^a_+(z_1) J^b_+(z_2)  & \sim \frac{2 \delta^{a,b}}{z_{12}^2} + \frac{1}{z_{12}} i \fabc J^c_+(z_2)~, 
\end{align}
where $z_{12}\equiv z_1-z_2$, and $\fabc$ are the structure constants of $\Le_8$ defined relative to the normalization where the roots have length squared $2$. 
Alternatively, the current correlation functions are determined by
\begin{align}
\label{eqn:corrplus}
\la J_+^a(z_1) J^b_{+}(z_2) \ra &= \frac{2 \delta^{a,b}}{z_{12}^2} ~,&
\la J_+^a(z_1) J^b_{+}(z_2) J_+^c(z_3)  \ra & = \frac{1}{z_{12} z_{23} z_{13}} i \cF^{abc}~,
\end{align}
where $\cF^{abc}=\fabd 2 \delta^{c,d}$. 

Turning this around, given the two- and three-point functions of any set of currents we can obtain the Cartan-Killing metric $\cG^{ab}$ and the structure constants $\fabc$ of the algebra as
\begin{align}
\label{eqn:algebra}
\cG^{ab} & \equiv z_{12}^2 \la J_+^a(z_1) J^b_{+}(z_2) \ra \, , \\
i\fabc & \equiv i \cF^{abd}\cG_{dc} =
z_{12} z_{23} z_{13} \la J_+^a(z_1) J^b_{+}(z_2) J_+^c(z_3)  \ra \cG_{dc} ~.
\end{align}

\section{Decompactification limit}
\label{sec:rinfty}
We now have the tools to analyze the symmetry enhancements of the decompactification limits. In this section, we focus on the case with zero Wilson line and the limit $R \to \infty$. Geometrically, the CHL string is obtained from the $(\text{E}_8 \times \text{E}_8) \rtimes \mathbb{Z}_2$ string compactified on $S^1$ with a holonomy for the $\mathbb{Z}_2$ exchange gauge symmetry along the circle direction, and therefore it is expected that by taking the decompactification limit the effect of the holonomy should disappear, thus giving back the ten dimensional $\text{E}_8 \times \text{E}_8$ heterotic theory. 
In this limit, the $(\Le_8)_2 \oplus \Lu(1)$ currents acquire a tower of approximately holomorphic light operators obtained by ``dressing'' the currents with vertex operators constructed from the compact boson:\footnote{Note that the dressing we introduce has winding mode number $\text{w}=0$, since all states with $\text{w}\neq 0$ have large conformal dimensions in the $R\to\infty$ limit.  For the same reason, the twisted sector does not contribute to light operators, since there $\text{w}\in \mathbb{Z}+\frac{1}{2}$.} 
\begin{align}
\label{eqn:jnp}
    \cJ^a_{+\text{n}}(z,\bar{z}) &= J^a_+(z) e^{i2\text{n}X^9(z,\bar{z})} \, , \, \text{n} \in \mathbb{Z} \,, \\
\label{eqn:jn9}    
    \cJ^9_{\text{n}}(z,\bar{z}) &=  J^9(z)e^{i2\text{n}X^9(z,\bar{z})}\, , \, \text{n} \in \mathbb{Z} \, , 
\end{align}
where the normal ordering of the operators is understood. These operators have weights $(\text{h},\bar{\text{h}})=\left( 1+ \frac{\text{n}^2}{R^2}, \frac{\text{n}^2}{R^2} \right) \to (1,0)$, and we refer to such operators as asymptotic currents. Note that the orbifold projection requires the momentum quantum number to be even.
 
Another set of asymptotic currents is constructed by dressing the antisymmetric combinations of the $(\Le_8)_1 \oplus (\Le_8)_1$ currents:
\begin{equation}
\label{eqn:jnm}
    \cJ^{\dot{a}}_{-\text{r}}(z,\bar{z}) = J^{\dot{a}}_-(z) e^{i2\text{r}X^9(z,\bar{z})}\, ,~ \, \text{r} \in \mathbb{Z}+\frac{1}{2}
\end{equation}
where 
\begin{align}
\label{eqn:antisymcur}
    J^{a}_-(z) =J^a_1(z)-J^a_2(z) \, , 
\end{align}
and in \eqref{eqn:jnm} the momentum $\text{n} = 2\text{r}$ surviving the projection is odd. Note that we assign half-integer labels to the $\cJ_{-\text{r}}$ currents, and we introduced a dotted index $\dot{a}$, also valued in the adjoint of $\Le_8$, to distinguish the $\cJ_{\pm}$ asymptotic currents in what follows.\footnote{The asymptotic currents are also charged with respect to the translation current $J^9$, which leads to the extension of the algebra generated by the $\cJ_{\pm}$.} 

We can now use \eqref{eqn:algebra} to obtain the Cartan-Killing metric and structure constants of the algebra of the $\cJ_\pm$ and $\cJ^9$ currents in the limit $\ep = \frac{1}{2 R^2} \to 0$.  To do so, we first note that the two- and three-point functions of the even and odd currents are
\begin{align}
\la J^a_{+}(z_1) J^b_{-}(z_2) \ra& = 0~, &
\la J_{\pm}^{a}(z_1) J_{\pm}^b(z_2) \ra &= \frac{2\delta^{a,b}}{z_{12}^2}~,
\end{align}
and
\begin{align}
\label{eqn:corr1}
\la J_{+}^{a}(z_1) J_{+}^b(z_2) J_{+}^c(z_3) \ra &=\la J_{-}^{a}(z_1) J_{-}^b(z_2) J_{+}^c(z_3) \ra= \frac{i \cF^{abc}}{z_{12} z_{23} z_{13}}~, \\
\label{eqn:corr2}
\la J_{-}^{a}(z_1) J_{+}^b(z_2) J_{+}^c(z_3) \ra &=\la J_{-}^{a}(z_1) J_{-}^b(z_2) J_{-}^c(z_3) \ra =  0~. 
\end{align}
These results follow 
since the $J^a_-$ are found to be in the adjoint representation of the $(\Le_8)_2$ and are odd under the exchange symmetry.  We also have the three-point function for the compact boson (the normal ordering of the exponentials is understood)
\begin{align}
\la e^{i\text{n}_1 X^9(z_1,\bar{z}_1)}e^{i\text{n}_2 X^9(z_2,\bar{z}_2)}e^{i\text{n}_3 X^9(z_3,\bar{z}_3)}\ra
= \frac{\delta_{n_1+n_2+n_3,0}}{|z_{12}|^{2(h_1+h_2-h_3)}|z_{23}|^{2(h_2+h_3-h_1)}|z_{13}|^{2(h_1+h_3-h_2)}}~,
\end{align}
where $h_i = \frac{\text{n}_i}{4R^2}$.

As far as the asymptotic current algebra is concerned, only the zero-momentum copy of the $\Lu(1)$ contributes \cite{Collazuol:2022jiy};  more precisely, including $\cJ^9_{\text{n}}$ with $\text{n}\neq 0$ would lead to an additional pole in the three-point functions that is inconsistent with the form of a current algebra.  Thus, we restrict to $\cJ^9_0\equiv\cJ^9$ in what follows.  Its correlation functions with insertions of $\cJ_{\pm}$ are fixed by the $\Lu(1)$ Ward identity.

Putting together these explicit correlation functions, we find that the non-trivial two-point functions are
\begin{align}
\label{eq:2pointchl}
\cGt^{a\text{m};b\text{n}} &:=z_{12}^2 \la \cJ^a_{+\text{m}}(z_1,\bar{z}_1) \cJ^b_{+\text{n}}(z_2,\bar{z}_2)\ra = 2\delta^{a,b}\delta^{\text{m}+\text{n},0}\left(1+ \GO (\ep \log|z_{12}|\right)) \, ,~\nonumber\\
\cGt^{\dot{a}\text{r};\dot{b}\text{s}} &:=z_{12}^2 \la\cJ^{\dot{a}}_{-\text{r}}(z_1,\bar{z}_1) \cJ^{\dot{b}}_{-\text{s}}(z_2,\bar{z}_2)\ra = 2\delta^{\dot{a},\dot{b}}\delta^{\text{r}+\text{s},0}\left(1+ \GO(\ep \log|z_{12}|\right)) \, , ~\nonumber\\
\cGt^{9,9} & :=z_{12}^2 \la \cJ^9_{n}(z_1,\bar{z}_1) \cJ^9_{n}(z_2,\bar{z}_2) \ra  = \frac{1}{R^2} =  2 \ep ~.
\end{align}
As expected, all anti-holomorphic dependence disappears in the $\ep \to 0$ limit.
Similarly, the non-zero three-point functions are
\begin{align}
\label{eq:3pointchl}
\cFt^{a\text{m};b\text{n};c\text{p}} &:= z_{12} z_{23} z_{13} \la \cJ^a_{+\text{m}}(z_1,\bar{z}_1) \cJ^b_{+\text{n}}(z_2,\bar{z}_2) \cJ^c_{+\text{p}}(z_3,\bar{z}_3) \ra
= 2 f^{abc} \delta^{\text{m}+\text{n}+\text{p},0}\left(1 + \GO(\ep \log|z_{ij}|)\right)~,
\nonumber\\
\cFt^{\dot{a}\text{r};\dot{b}\text{s};c\text{p}} &:= z_{12} z_{23} z_{13} \la \cJ^{\dot{a}}_{-\text{r}}(z_1,\bar{z}_1) \cJ^{\dot{b}}_{-\text{s}}(z_2,\bar{z}_2) \cJ^c_{+\text{p} }(z_3,\bar{z}_3)\ra
= 2 f^{\dot{a}\dot{b}c} \delta^{\text{r}+\text{s}+\text{p},0}\left(1 + \GO(\ep \log|z_{ij}|)\right)~,
\nonumber\\
\cFt^{a\text{m};b\text{n};9} &:= z_{12} z_{23} z_{13} \la \cJ^a_{+\text{m}}(z_1,\bar{z}_1) \cJ^b_{+\text{n}}(z_2,\bar{z}_2) \cJ^9(z_3)\ra
= -2i \frac{2\text{m}}{R^2\sqrt{2}} \delta^{a,b} \delta^{\text{m}+\text{n},0}\left(1 + \GO(\ep \log|z_{ij}|)\right)~,
\nonumber\\
\cFt^{\dot{a}\text{r};\dot{b}\text{s};9} &:= z_{12} z_{23} z_{13} \la \cJ^{\dot{a}}_{-\text{r}}(z_1,\bar{z}_1) \cJ^{\dot{b}}_{-\text{s}}(z_2,\bar{z}_2) \cJ^9(z_3)\ra
= -2i \frac{2\text{r}}{R^2\sqrt{2}} \delta^{a,b} \delta^{\text{r}+\text{s},0}\left(1 + \GO(\ep \log|z_{ij}|)\right)~.
\end{align}
Thus, the asymptotic currents give rise to an algebra whose non-trivial components of the Cartan-Killing metric are
\begin{align}
\cGt^{a\text{m};b\text{n}}  = 2\delta^{a,b}\delta^{\text{m}+\text{n},0}~, \quad
\cGt^{\dot{a}\text{r};\dot{b}\text{s}}  = 2\delta^{\dot{a},\dot{b}}\delta^{\text{r}+\text{s},0}~,&
\end{align}
and whose structure constants are given by
\begin{align}
    \ft^{a\text{m};b\text{n}}_{~~~~~~~c\text{p}}  = f^{ab}_{~~~c} \delta^{\text{m}+\text{n}+\text{p},0}~,\quad  \ft^{\dot{a}\text{r};\dot{b}\text{s}}_{~~~~~~c\text{p}} & = f^{\dot{a}\dot{b}}_{~~~c} \delta^{\text{r}+\text{s}+\text{p},0}~, \quad
    \ft^{a\text{m};b\text{n}}_{~~~~~~9} = -2i\text{m}\sqrt{2}\delta^{a,b}\delta^{\text{m}+\text{n},0}~,  \nonumber\\
    \ft^{\dot{a}\text{r};\dot{b}\text{s}}_{~~~~~~9} & = -2i\text{r}\sqrt{2}\delta^{\dot{a},\dot{b}}\delta^{\text{r}+\text{s},0}~.
\end{align}
Because $\text{r} \in \mathbbm{Z} + \frac{1}{2}$, they correspond to the twisted version of the algebra $(\Le_9 \oplus \Le_9) / \sim$ (see for instance \cite{cmp/1104159538} and \cite{Goddard:1986bp}), which we will denote as $(\Le_9 \oplus \Le_9)_{\text{tw}} / \sim$. The $\sim$ stands for the identification of the central extensions for the two $\Le_9$, corresponding to $\cJ^9$. 

\subsection*{Comparison with heterotic decompactification: the twist}
We will now compare this with the asymptotic currents in the decompactification limit of the full heterotic theory from nine to ten dimensions. The generators of the $(\Le_9 \oplus \Le_9) / \sim$ in that case are given by
\begin{align}
    \cK^a_{+\text{n}}(z,\bar{z}) &= J^a_+(z) e^{i\text{n}X^9(z,\bar{z})} \, , \, \text{n} \in \mathbb{Z} \,, \\
\label{eqn:knphet}
    \cK^{\dot{a}}_{-\text{r}}(z,\bar{z}) &= J^{\dot{a}}_-(z) e^{i\text{r}X^9(z,\bar{z})} \, , \, \text{r} \in \mathbb{Z} \,, \\    
\label{eqn:kn9het}    
    \cK^9(z) &=  J^9(z)\, . 
\end{align}
These currents have both even and odd momentum, and they are labelled by integers. In this case, following the same procedure, one obtains
\begin{align}
\cG^{a\text{m};b\text{n}} & = 2\delta^{a,b}\delta^{\text{m}+\text{n},0}~,&
\cG^{\dot{a}\text{r};\dot{b}\text{s}} & = 2\delta^{\dot{a},\dot{b}}\delta^{\text{r}+\text{s},0}~,&
f^{a\text{m};b\text{n}}_{~~~~~~~c\text{p}} & = f^{ab}_{~~~c} \delta^{\text{m}+\text{n}+\text{p},0}~,
\nonumber\\
f^{\dot{a}\text{r};\dot{b}\text{s}}_{~~~~~~~c\text{p}} & =
f^{\dot{a}\dot{b}}_{~~~c} \delta^{\text{r}+\text{s}+\text{p},0}~,&
f^{a\text{m};b\text{n}}_{~~~~~~9} & = -i\text{m}\sqrt{2}\delta^{a,b}\delta^{\text{m}+\text{n},0}~, &
f^{\dot{a}\text{r};\dot{b}\text{s}}_{~~~~~~9} & = -i\text{r}\sqrt{2}\delta^{\dot{a},\dot{b}}\delta^{\text{r}+\text{s},0}~.
\end{align}
In the heterotic case the structure constants involving the $\cK^9(z)$ are smaller by a factor $2$ than in the CHL case. This can be absorbed into a rescaling of $\cK^9(z)$: we set $\cK^{9'}(z) = 2 \cK^9(z)$, which leads to
\begin{align}
\cGt^{9',9'} = \frac{4}{R^2} = 8 \ep~, 
\end{align}
so that
\begin{align}
\cFt^{a\text{m};b\text{n};9'} &= 2 \cFt^{a\text{m};b\text{n};9}~,
\end{align}
and
\begin{align}
f^{a\text{m};b\text{n}}_{~~~~~~~9'} = \frac{1}{2} f^{a\text{m};b\text{n}}_{~~~~~~~9}~.
\end{align}
Similarly $f^{\dot{a}\text{r};\dot{b}\text{s}}_{~~~~~~9'} = \frac{1}{2} f^{\dot{a}\text{r};\dot{b}\text{s}}_{~~~~~~9}$ .

The rescaling has a simple interpretation: the CHL orbifold naturally gives us a geometry with radius $R_{\text{CHL}} = R/2$.  On the other hand, the fact that in \eqref{eqn:jnm} $\text{r} \in \mathbb{Z}+\frac{1}{2}$ while in \eqref{eqn:knphet} $\text{r} \in \mathbb{Z}$ is a structural difference: the CHL algebra is obtained through a twisting of the heterotic one by an outer automorphism of $\text{E}_8 \times \text{E}_8$, and the decompactification retains this feature as $R\to \infty$. This is our key result.

While we obtained our result by setting $a=0$, it can be easily generalized to the case of generic Wilson line because in the $R\to \infty$ limit its contribution to the conformal dimension of the $(\Le_9 \oplus \Le_9)_{\text{tw}} / \sim$ currents vanishes:  $(\text{h},\bar{\text{h}})= \left( 1 + \frac{(\text{n}-\alpha \cdot a)^2}{4R^2}, \frac{(\text{n}-\alpha \cdot a)^2}{4R^2} \right) \to (1,0)$ for each root $\alpha \in \Gamma_8$ (see \cite{Collazuol:2022oey} for the equivalent in the full heterotic picture), so that the affine enhancement is the same.

\section{Decompactification limit, dual frame}
\label{sec:rzero}
In this section, we consider the T-dual decompactification limit, which according to \eqref{eqn:a} is given by $R \to 0$ and $a =0$. We do this for several reasons. First, we want to directly check the presence of $(\Le_9 \oplus \Le_9)_{\text{tw}} / \sim$ . This is not trivial because in this limit the asymptotic currents arise in both sectors of the orbifold, leading to significant differences with respect to the heterotic decompactification and its T-dual. In addition, the explicit realization that we will obtain may be useful for understanding how twisted and untwisted currents concur to give level two enhancements at finite distance in moduli space.

To understand the asymptotic currents in the $R \to 0$ limit, we will need to describe the twisted Hilbert space in the orbifold CFT. This is most easily accomplished by first constructing the exchange orbifold by $\mathcal{E}$ of the ten dimensional heterotic theory. We know that in this case we recover an isomorphic theory \cite{Forgacs:1988iw,Chaudhuri:1995bf}.

\subsection{The exchange orbifold in ten dimensions}
Following \cite{Forgacs:1988iw}, one can decompose the $\text{E}_8 \times \text{E}_8$ internal holomorphic CFT with $c=16$ of the heterotic string, with energy-momentum tensor $T_{\text{E}_8 \times \text{E}_8}(z)$, in terms of the two commuting factors
\begin{equation}
\label{eqn:split}
    T_{\text{E}_8 \times \text{E}_8}(z) = T_{\text{E}_{8,2}}(z)+T_{\text{Ising}}(z) \, ,
\end{equation}
where the $\text{E}_8$ Kac-Moody algebra is realized at level 2 on the worldsheet by the modes of the $\{ J^a_+(z) \}$ and has central charge $c_2=\frac{31}{2}$. There are three unitary, highest weight, integrable representations $\lambda$ of the $(\widehat{\Le}_8)_2$ sector, labeled by the $\Le_8$ irreducible representation
\cite{Kac_1990,DiFrancesco:1997nk} :
\begin{align}
    \lambda & =\boldsymbol{1} \, ,  &\text{h}_{\boldsymbol{1}} &=0 \, , \nonumber \\
    \lambda & =\boldsymbol{248}\, ,  &\text{h}_{\boldsymbol{248}} &=  \frac{15}{16}\, , \nonumber \\
    \label{eqn:prime8}
    \lambda& =\boldsymbol{3875} \, ,   &\text{h}_{\boldsymbol{3875}}&  = \frac{3}{2} \, .
\end{align}
The familiar primary fields of the Ising CFT have conformal weights
\begin{align}
\label{eqn:primising}
    \text{h} =0 \, , \quad \text{h} =\frac{1}{16} \, , \quad    
    \text{h}  =\frac{1}{2} \, , 
\end{align}
corresponding respectively to the vacuum, the (holomorphic) spin field and the (holomorphic) Majorana-Weyl fermion. 

From \eqref{eqn:split}, it is natural to express the partition function $Z_{\GE_8 \times \GE_8}(\tau)=Z_{\GE_8}(\tau)^2$ in terms of the characters of $(\widehat{\Le}_8)_2$ \cite{Forgacs:1988iw}, based on the primaries described in \eqref{eqn:prime8}:
\begin{align}
\cX^{\GE_{8,2}}_{\rep{1}} & = q^{-31/48} \left( 1 + 248 q + 31\,124 q^2 + 871\,627 q^3 + \GO(q^4) \right)~, \nonumber\\
\cX^{\GE_{8,2}}_{\rep{248}} & = q^{14/48} \left(248 + 34\,504 q + 1\,022\,752 q^2 + \GO(q^3) \right)~,\nonumber\\
\cX^{\GE_{8,2}}_{\rep{3875}} & = q^{41/48} 31\left(125 + 5863 q + 116\,899q^3 + \GO(q^4) \right)~,
\end{align}
and of the Ising CFT ones, associated to the primaries in \eqref{eqn:primising}:
\begin{align}
\cX^{\text{I}}_0 & = q^{-1/48} \left(1+q^2+q^3 + 2 q^4 +\GO(q^5) \right)~,\nonumber\\
\cX^{\text{I}}_{1/2} & = q^{-1/48+1/2} \left(1 + q + q^2 + q^3 +\GO(q^4) \right)~,\nonumber\\
\cX^{\text{I}}_{1/16} & = q^{2/48} \left(1 + q + q^2 + 2q^3 + \GO(q^4) \right)~.
\end{align}
As usual, $q$ denotes the modular parameter $q=e^{2\pi i \tau}$. The result of~\cite{Forgacs:1988iw} is that 
\begin{align}
\label{eqn:ze8e8}
Z_{\GE_8}(\tau)^2 = \cX^\text{I}_0 \cX^{\GE_{8,2}}_{\rep{1}} + \cX^\text{I}_{1/2} \cX^{\GE_{8,2}}_{\rep{3875}} + \cX^\text{I}_{1/16} \cX^{\GE_{8,2}}_{\rep{248}}~,
\end{align}
where the first two terms are even under the permutation symmetry, while the last one is odd.\footnote{This decomposition has also been recently reviewed in the context of classification of holomorphic CFTs with $c\le 16$~\cite{BoyleSmith:2023xkd}.}

To connect this presentation with the previous discussion, the $(\widehat{\Le}_8)_2$ current algebra is generated by the $J^{a}_{+}$, while the $J^a_{-}$ are in the adjoint representation and correspond to the last ($\mathcal{E}$-odd) term in \eqref{eqn:ze8e8}.

To obtain the $\mathcal{E}$ exchange orbifold we can follow~\cite{Klemm:1990df}, which gave a general construction for the partition function of a cyclic permutation orbifold.\footnote{This was then generalized in, e.g.~\cite{Fuchs:1991vu,Borisov:1997nc}.}  Rather than discuss the general situation, we focus on the case with a $\Z_2$ permutation.  Let $\cC$ be a CFT with a modular-invariant partition function $Z_{\cC}$.  Then the product theory has partition function
\begin{align}
Z_{0}(\tau,\taub) & = Z_{\cC}(\tau,\taub)^2~.
\end{align}
By taking modular orbits, one arrives at the $\cC\otimes\cC/\Z_2$ partition function:\footnote{See also \cite{Albert:2022gcs} for a modern presentation and discussion of possible anomalies.}
\begin{align}
Z_{\text{orb}} & = \frac{1}{2} Z_0(\tau,\taub) + \frac{1}{2} Z_{\cC}(2\tau,2\taub) + \frac{1}{2} Z_{\cC}\left(\ff{\tau}{2},\ff{\taub}{2}\right) + \frac{1}{2} Z_{\cC}\left(\ff{\tau+1}{2},\ff{\taub+1}{2}\right)~.
\end{align}
We cannot apply their construction verbatim to the partition function for $\GE_8$ because the latter is not modular--invariant, and instead it picks up a phase under the T transformation $\tau \to \tau +1$:
\begin{align}
Z_{\GE_8}(\tau+1) &= e^{-2\pi i /3} Z_{\GE_8}(\tau)~, &
Z_{\GE_8}(-1/\tau) & = Z_{\GE_8}(\tau)~.
\end{align}
The phase factor just arises from the overall factor of $q^{-c/24}=q^{-1/3}$ in $Z_{\text{E}_8}(\tau)$ .

However, we can attempt to construct a partition function that would have the same covariance. To do this in the most obvious fashion, write the full modular-invariant partition function:
\begin{align}
Z_0(\tau,\taub)& = Z_{\GE_8}(\tau)^2 \Zt(\tau,\taub)~,
\end{align}
where the second factor satisfies
\begin{align}
\Zt(\tau+1,\taub+1) &= e^{4\pi i /3} \Zt(\tau,\taub)~, &
\Zt(-1/\tau,-1/\taub) & = \Zt(\tau,\taub)~.
\end{align}
Taking invariant states, we obtain
\begin{align}
Z_{\text{inv}} (\tau,\taub) & =  \frac{1}{2} \left(Z_{\GE_8}(\tau)^2 + Z_{\GE_8}(2\tau)\right) \Zt(\tau,\taub)~.
\end{align}
This is invariant under the T transformation $\tau\to\tau+1$ because $Z_{\GE_8}(\tau)^2$ and $Z_{\GE_8}(2\tau)$ come with the same overall factor of $q^{-2\times 8/24}$.  While the first term is also invariant under the S transformation $\tau \to - \frac{1}{\tau}$,  the second one is not, but by taking the modular orbit and using S invariance and T covariance of $Z_{\text{E}_8}(\tau)$ we find a candidate for the modular-invariant partition function:
\begin{align}
Z_{\text{orb}} & =  \frac{1}{2} \left(Z_{\GE_8}(\tau)^2 + Z_{\GE_8}(2\tau)\right) \Zt(\tau,\taub) \nonumber\\
&\qquad
+\frac{1}{2} Z_{\GE_8}(\ff{\tau}{2}) \Zt({\tau},{\taub})
+\frac{1}{2} Z_{\GE_8}(\ff{\tau+1}{2}) e^{4\pi i/3} \Zt({\tau},{\taub})~ \nonumber\\
& = Z'_{\text{orb}}(\tau) \Zt(\tau,\taub)~,
\end{align}
where
\begin{align}
\label{eqn:zorb}
Z'_{\text{orb}}(\tau) &= \frac{1}{2} \left( Z_{\GE_8}(\tau)^2+Z_{\GE_8}(2\tau)+Z_{\GE_8}(\ff{\tau}{2})+e^{4\pi i/3}Z_{\GE_8}(\ff{\tau+1}{2})\right) \\ & = Z_{\GE_8}(\tau)^2~ \nonumber.
\end{align}
The last equality is the non-trivial statement~\cite{Forgacs:1988iw} that this orbifold is equivalent to the original theory. This can be seen explicitly by writing each term in $Z'_{\text{orb}}(\tau)$ in terms of the $(\widehat{\Le}_8)_2$ and Ising characters
\begin{align}
\label{eqn:sectors}
Z_{\GE_8}(\tau)^2 & = + \cX^{\text{I}}_{0} \cX^{\GE_{8,2}}_{\rep{1}} +\cX^{\text{I}}_{1/2} \cX^{\GE_{8,2}}_{\rep{3875}}
+\cX^{\text{I}}_{1/16}\cX^{\GE_{8,2}}_{\rep{248}} ~,\nonumber\\
Z_{\GE_8}(2\tau) & = + \cX^{\text{I}}_{0} \cX^{\GE_{8,2}}_{\rep{1}} + \cX^{\text{I}}_{1/2} \cX^{\GE_{8,2}}_{\rep{3875}}
-\cX^{\text{I}}_{1/16} \cX^{\GE_{8,2}}_{\rep{248}}~,\nonumber\\
Z_{\GE_8}(\ff{\tau}{2})  & =+ \cX^{\text{I}}_{1/2} \cX^{\GE_{8,2}}_{\rep{1}} +\cX^{\text{I}}_{0} \cX^{\GE_{8,2}}_{\rep{3875}}
+\cX^{\text{I}}_{1/16}\cX^{\GE_{8,2}}_{\rep{248}}~,\nonumber\\
e^{4\pi i} Z_{\GE_8}(\ff{\tau+1}{2}) & = -\cX^{\text{I}}_{1/2} \cX^{\GE_{8,2}}_{\rep{1}} -\cX^{\text{I}}_{0} \cX^{\GE_{8,2}}_{\rep{3875}}
+\cX^{\text{I}}_{1/16} \cX^{\GE_{8,2}}_{\rep{248}}~,
\end{align}
so that indeed the sum correctly reproduces $2 Z_{\GE_8}(\tau)^2 $ as expanded in \eqref{eqn:ze8e8}.

The terms appearing in \eqref{eqn:zorb} can be interpreted in terms of a projection in the untwisted and twisted Hilbert spaces $\cH_{\text{ut}}$ and $\cH_{\text{t}}$ respectively as \begin{align}
Z_{\text{ut}}(\tau)& =\Tr_{\cH_{\text{ut}}} \left\{ (1+\mathcal{E}) q^{L_0-c/24}\right\} = Z_{\GE_8}(\tau)^2+Z_{\GE_8}(2\tau)~, \nonumber \\
\label{eqn:ztw}
Z_{\text{t}}(\tau)& =\Tr_{\cH_{\text{t}}} \left\{ (1+\mathcal{E}) q^{L_0-c/24}\right\} = Z_{\GE_8}(\ff{\tau}{2})+e^{4\pi i/3}Z_{\GE_8}(\ff{\tau+1}{2})~.
\end{align}
In particular, focusing on \eqref{eqn:ztw}, we can characterize the $\mathcal{E}$-even and -odd states by expanding the two terms separately.  For our purposes, it suffices to examine the leading terms in the expansions:
\begin{align}
\label{eq:twistedsectors}
Z_{\GE_8}(\ff{\tau}{2}) &= q^{-1/6} \left(+1 +248 q^{1/2} + 4\,124 q + 34\,752 q^{3/2}+213\,126 q^2 + \GO(q^{5/2})\right)~,
\nonumber\\
e^{4\pi i} Z_{\GE_8}(\ff{\tau+1}{2}) & = q^{-1/6} \left(-1 +248 q^{1/2} - 4\,124 q + 34\,752 q^{3/2}-213\,126 q^2  + \GO(q^{5/2})\right)~.
\end{align}
Writing $q^{-1/6} = q^{-2/3} q^{1/2}$ , we learn that the twisted ground state is $\mathcal{E}$-odd and has $\text{h}=\frac{1}{2}$. We denote the corresponding twist field as $\Lambda(z)$. 
 The additional currents can be read off directly from the $Z_{\text{t}}(\tau)$ as coming from the $\text{h}=1$ primaries in $\cX^{\text{I}}_{1/16} \cX^{\GE_{8,2}}_{\rep{248}}$ of the last line in~\eqref{eqn:sectors}, \footnote{This could have equivalently been argued from the $q^{1/3}$ term in \eqref{eq:twistedsectors}.
} which are the twisted version of the $J^a_-(z)$ in \eqref{eqn:antisymcur}. While $J^a_-$ are $\mathcal{E}$-odd, we have the OPE 
\begin{equation}
\label{eqn:jl}
    J^a_-(z_1) \Lambda(z_2) \sim \frac{K^a(z_2)}{z_{12}^{1/2}} \, ,
\end{equation}
where the $K^a(z)$ are the $\mathcal{E}$-even primary operators associated to the $\cX^{\text{I}}_{1/16} \cX^{\GE_{8,2}}_{\rep{248}}$ characters in the twisted Hilbert space. Since $\Lambda(z)$ is a singlet under  $(\widehat{\Le}_8)_2$, the $K^{a}(z)$ are in the adjoint representation of the algebra:
\begin{equation}
    J^a_+(z_1)K^b(z_2) \sim \frac{i \fabc K^{c}(z_2)}{z_{12}} \, ,
\end{equation}
with $\fabc$ the $\Le_8$ structure constants.
The conformal and $(\widehat{\Le}_8)_2$ Ward identities imply that 
\begin{align}
    \la J^a_+(z_1) K^{b}(z_2) K^{c}(z_3) \ra  = i \fabd \la K^{d}(z_2) K^{c}(z_3)\ra \frac{z_{23}}{z_{12}z_{13}} \, .
\end{align}
We normalize the $K^a(z)$ as
\begin{align}
    \la K^a(z_1) K^b(z_2) \ra = \frac{\cG^{ab}}{z_{12}^2} \, ,
\end{align}
with $\cG^{ab} = 2 \delta^{a,b}$ as the Cartan-Killing metric of $(\Le_{8})_2$ (this just amounts to choosing accordingly the normalization of the twist operator) and obtain 
\begin{align}
    \la J_{+}^{a}(z_1) K^b(z_2) K^c(z_3) \ra= \frac{i \cF^{abc}}{z_{12} z_{23} z_{13}}~,
\end{align}
with the $\cF^{abc}$ as in \eqref{eqn:corr1}.

Moreover, the quantum symmetry of the orbifold, which assigns charge $+1$ to the untwisted and $-1$ to the twisted sector, implies that
\begin{align}
    \la K^{a}(z_1) J^{b}_+(z_2) J^c_+(z_3) \ra & =0 \, ,
    \\
    \la K^{a}(z_1) K^{b}(z_2) K^c(z_3) \ra & =0 \, ,
\end{align}
so that we recover the full description of the algebra $(\Le_8)_1 \oplus (\Le_8)_1$~, but now we also have a detailed understanding of the orbifold action on the currents, as well as the structure of the twisted sector.

\subsection{Decompactification limit in the dual $R \to 0$ frame}
Having reviewed the pure exchange orbifold $\mathcal{E}$, we now combine it with the shift $\mathcal{S}$ on the circle, focusing on the $R \to 0$ limit in the absence of Wilson lines. In this case, the holomorphic currents are dressed with winding modes, and the orbifold pairs the $J^a_+$ with integer winding in the untwisted sector and $K^a$ with half-integer winding in the twisted sector. This is what we expect by T-duality \eqref{eqn:a} applied to the asymptotic currents found in Section \ref{sec:rinfty}, with the correspondence being
\begin{align}
\label{eqn:jnpdual}
    \cJ^a_{+\text{n}}(z,\bar{z}), \, \text{n} \in \mathbb{Z} & \leftrightarrow J^a_+(z) e^{i\frac{\text{w}}{\sqrt{2}}R^2\left(X_L^9(z) - X^9_R(\bar{z})\right)} \, , \, \text{w} \in \mathbb{Z} \,, \\
\label{eqn:jn9dual}    
     \cJ^{\dot{a}}_{-\text{r}}(z, \bar{z}) , \, \text{r} \in \mathbb{Z}+\frac{1}{2} & \leftrightarrow  K^{\dot{a}}(z) e^{i\frac{\text{w}}{\sqrt{2}}R^2\left(X_L^9(z) - X^9_R(\bar{z})\right)} \, , \, \text{w} \in \mathbb{Z}+\frac{1}{2} \, . 
\end{align}
In the $R\to \infty$ limit, the twisting in the asymptotic current algebra can be traced back to the different moding of the $ J_+$ and $J_-$ towers. We now see that in the T-dual frame this is a consequence of the fact that in the two sectors the winding has either integer or half-integer values.

\section{Conclusions}
\label{sec:conclusion}
In this paper we focused on one type of infinite distance limit in the rank $9$ component of the moduli space in nine dimensions: the decompactification of the perturbative CHL string to ten dimensions, which is characterized by a twisted affine algebra. Our result can be generalized to the case of eight-dimensional compactifications with rank reduction of CHL type when one decompactifies the cycle carrying the holonomy. Thus, it is a non-trivial prediction to be reproduced in a dual M/F-theory description in the presence of frozen singularities on K3 in the appropriate infinite distance limit. In lower dimensions there are also other possibilities; for example, in the case of the CHL string on $S^1$, one could decompactify the $S^1$ without the holonomy recovering an affine, non-twisted algebra (from the Type II point of view, see \cite{Cvetic:2022uuu}). 

Another possible avenue would be to understand the algebras arising in infinite distance limits in the case of Calabi-Yau compactifications of the heterotic theory (which in the K3 case is dual to F-theory on elliptic Calabi-Yau threefolds, whose degenerations at infinite distance have been recently considered in \cite{Alvarez-Garcia:2023gdd,Alvarez-Garcia:2023qqj}). However, even in the heterotic frame, this would require an understanding of both space-time and worldsheet non-perturbative effects.   It may be possible and instructive to generalize our methods to special loci where the K3 is realized as an orbifold of $T^4$, and we hope that the lessons learned from those examples may be useful for understanding the more general situation.

While we focused on decompactification limits that preserve the number of supercharges, we could also consider the case of nine-dimensional non-supersymmetric heterotic string, in which case we would expect to find both an asymptotic current algebra and supersymmetry enhancement, at least at string tree-level.  It would then be interesting to reconsider these structures in light of the non-zero vacuum energy generated by string loops.

\subsection*{Acknowledgements}
VC's work is supported by CNRS-Imperial College PhD fellowship.  
IVM's work is partially supported by the Humboldt Research Award and the Jean d'Alembert Program at the Universit{\'e} Paris--Saclay hosted at IPhT, as well as the Educational Leave program at James Madison University.  
Part of this work was carried out while IVM was visiting  the Albert Einstein Institute (Max Planck Institute for Gravitational Physics), LPTHE at Sorbonne Universit{\'e}, as well as the Mathematical Physics group at Universit{\"a}t Wien, and he thanks all three institutions for hospitality.  We are grateful to P.~Cheng, G.~Di Ubaldo, B.~Fraiman, S.~Fredenhagen, S.~Georgescu, A.~Kleinschmidt, C.~A.~Nu{\~n}ez and H.~Parra De Freitas for valuable discussions.
We especially thank M.~Gra{\~n}a for her contributions and initial collaboration in the project.

\bibliographystyle{JHEP}
\bibliography{biblio}

\end{document}